# Research on Wearable Technologies for Learning: A Systematic Review

Sharon Lynn Chu[1], Brittany M. Garcia[2], Neha Rani[3(✉)]
[1,3]University of Florida, Gainesville, FL, USA. [2]Texas A&M University, College Station, TX, USA.
slchu@ufl.edu// brinni@tamu.edu// neharani@ufl.edu

**ABSTRACT:** A good amount of research has explored the use of wearables for educational or learning purposes. We have now reached a point when much literature can be found on that topic, but few attempts have been made to make sense of that literature from a holistic perspective. This paper presents a systematic review of the literature on wearables for learning. Literature was sourced from conferences and journals pertaining to technology and education, and through an ad hoc search. Our review focuses on identifying the ways that wearables have been used to support learning and provides perspectives on that issue from a historical dimension, and with regards to the types of wearables used, the populations targeted, and the settings addressed. Seven different ways of how wearables have been used to support learning were identified. We propose a framework identifying five main components that have been addressed in existing research on how wearables can support learning and present our interpretations of unaddressed research directions based on our review results.

## 1. Introduction

The desktop-bound paradigm of user interaction with technologies is rapidly losing its mainstream status to make way for a mode of interaction that is marked by mobility, persistence, and ubiquity. While mobile technologies kickstarted this new paradigm, wearable technologies are a class of devices where the benefits of this new mode of interaction can be truly evident. Research has explored the use of wearable devices for a variety of purposes. Some of the primary use cases have included, for example, medicine, healthcare and wellbeing, business, and military where the sensor capabilities of wearables enable useful tracking features. However, a good amount of work has also explored the potential of wearable devices in the domain of education. Although work in that area is still relatively scarce compared to wearables for health, there are now enough contributions to the literature to warrant a systematic review. Section II provides the context of such a review and make explicit the need for it.

The overarching goal of this paper is to provide the reader with a picture of the literature on educational wearables, assess its current status, and identify potential directions for future research. The results would be particularly useful for researchers who are new to the topic of wearables for learning, and practitioners (designers and educators) interested in understanding how the research community has engaged the issue thus far. The main research question that we addressed in our review is: *How have wearables been used to support learning in existing research?* We study this question with respect to: (a) the prevalence of research on wearables for learning over time; (b) the types of wearables used; (c) the populations targeted; and (d) the types of learning settings that the research is situated in.

## 2. Related Work

Work on wearable technologies for learning, or *educational wearables*, has advanced with few attempts at integration. Many reviews on wearables exist, but they address either wearables at a general level (i.e., aspects of wearables that are independent of application domain) (e.g., Berglund, Duvall, and Dunne (2016); Kumari, Mathew, and Syal (2017); Liew, Wah, Shuja, Daghighi, et al. (2015)), or wearables in the areas of healthcare (Baig, Gholamhosseini, and Connolly (2013); Pantelopoulos and Bourbakis (2008); Wang, Mintchev, et al. (2013)), assistive technologies (e.g., Dakopoulos and Bourbakis (2009); Tapu, Mocanu, and Tapu (2014)), or security (Blasco, Chen, Tapiador, and Peris-Lopez (2016); Sundararajan, Sarwat, and Pons (2019)). Our literature search uncovered 10 survey or review papers that can be considered as being related to wearables for learning.

The earliest review is a report published as part of the JISC Technology and Standards Watch Report in 2003 (De Freitas and Levene, 2003). It describes wearable devices, provides examples of wearables available at the time, and describes some case studies of how wearable technologies and mobile devices had been used in higher education settings. The report concludes with considerations, such as battery life, that need to be taken for the use of wearable and mobile devices in education. The next relevant survey paper authored by Petrovic (2014) in the journal of ICT Management, focuses on analyzing how some selected work have applied the use of wearables in education, and thereafter proposes two application models for how smart glasses and smartwatches can be combined in e-learning.

In 2015, there were 3 survey papers related to wearables for learning. Sapargaliyev (2015b) reviewed some work on how wearables have been used to support teaching and learning, but mainly used *GoogleGlass*. Sapargaliyev (2015a) published another 4-page paper reviewing work on wearables used in learning, this time with a broader scope. One conclusion from the paper is that "very little was found in the literature on the question of the use of wearables for large-scale projects". The last 2015 survey paper is by Borthwick, Anderson, Finsness, and Foulger (2015) in the Journal of Digital Learning in Teacher Education. The focus of the paper was to review the value and potential drawbacks of using wearable technologies in education. The authors identified some key themes for both value (e.g., student engagement, universal design for learning) and drawbacks (e.g., student safety, security, and privacy), highlighting example work supporting each theme. The paper wraps up with a call to action for researchers and educators to think of the theory of change that wearable technologies bring with respect to learning, and for more resources to be allocated to this topic.

In a 2018, Attallah and Ilagure (2018) published a survey that first describes some wearables available at the time (e.g., *GoogleGlass*, *Oculus Rift*, *Muse*), and then focuses on discussing the challenges associated with the use of wearable computing in education. Some limitations highlighted include distraction to students, cost, usability and fear of the technology, and the requirement of most wearables to be teetered to smartphones.

Lee and Shapiro (2019) conducted a survey that perhaps comes the closest to the review presented in this paper. Based on a review of a number of wearable technology investigations and projects, they identified the forms of support that wearables can provide for learning as including: (i) the promotion of personal expression; (ii) the integration of digital information into social interactions; (iii) the support of educative role-play; (iv) the provision of just-in-time notification in a learning environment; and (v) the production of records of bodily experience for later inspection, reflection, and interpretation.

Finally, in 2020, Havard and Podsiad (2020) conducted a meta-analysis of the effect sizes found in quantitative wearables for learning research. Their analysis included 12 studies. They also coded for various aspects of these studies, including the types of wearables used, and the pedagogical strategies used. They found that the majority of the studies (7 out of 12) used head-mounted displays, followed by fitness trackers and smartwatches. They classified the types of learning outcomes as being of a cognitive, affective, psychomotor, and motivational nature, with an overall weighted mean effect size for study outcomes of .6373 (medium effect).

The survey that we present in this paper is distinct from the previous surveys related to wearables for learning in that it focuses on how wearables have been used for learning, it is systematic in nature and more comprehensive, and covers a longer time period. Figure I show the distinctions between our survey in this paper and previous surveys on wearables for learning.

## 3. Obtaining the Paper Set

We describe the steps that we took to search for and review relevant papers below. The process is illustrated in Figure 2.

*A. Paper Search Process*

Three approaches were used to search for relevant papers:

(i) a researcher went through the entire proceedings/issues of 8 selected conferences and 5 selected journals for the last 13 years (2007 to April 2020) and identified potentially relevant papers by reading the paper titles and skimming the paper abstracts. The conferences and journals were selected because of the likelihood that they may include wearables and education. The list of selected conferences were as follows (Table I has full names of acronyms): CHI; SIGSCE; UbiComp; ISWC; MobileHCI; ACM International Conference on Interactive, Mobile, Wearable and Ubiquitous Technologies (IMWUT); ACM Annual Conference on Innovation and Technology in Computer Science Education (ITCSE); ACM International Conference on Advances in Mobile Computing and Multimedia (MoMM). The list of selected journals were as follows: Computers in Human Behavior; Computers and Education; IEEE Transactions on Education; IEEE Transactions on Learning Technologies; Learning, Media and Technology.

| Papers Aspect | De Freitas and Levene | Petrovic | Sapargaliyev | Sapargaliyev | Borthwick et al. | Attallah and Ilagure | Lee and Shapiro | Havard and Podsiad | Our Survey |
|---|---|---|---|---|---|---|---|---|---|
| Year | 2013 | 2014 | 2015 | 2015 | 2015 | 2018 | 2019 | 2020 | 2020 |
| Focus | How wearables combined with mobile devices have been used in higher education | Application models for how smart glasses and smart-watches can be combined in e-learning | Use of Google-Glass to support teaching and learning | Predict the possible barriers and problems in using new wearables in the classroom | Review the value and potential drawbacks of using wearable technologies in education | Challenges with the use of wearable computing in education | Identify the forms of support that wearables provide for learning | Effect sizes of learning outcomes for wearables | How wearables have been used to support learning |
| Systematic? | No | No | No | No | No | No | No | Yes | Yes |
| Methodology | Case studies | Case studies | Case studies | Case studies | Case studies | Case studies | Device analysis | Meta-analysis | Content analysis |

Fig. 1. Comparison of surveys on wearable for learning

(ii)  a search was performed using the Google Scholar search engine with combinations of the following search terms: 'wearable', 'wearables', 'wearable computing', 'learning', 'education', 'smart glasses', 'e-textile', 'smartwatch', 'smartwatches', 'wristbands', and 'smart jewelry'. Thus, a search phrase was for example "wearables learning". The search results pages for each search terms combination were reviewed until results began to appear irrelevant or repetitive. We note that searches through this approach was not limited by year of publication; and

(i)  the researcher reviewed the list of references of the papers found from the first and second approaches, as well as papers that cited those papers using the Google Scholar 'cited by' function. As for the previous approach, searches here were not limited by publication year.

A total of 349 papers were collected through the 3 approaches described above. From reviewing selected conference and journal proceedings, 203 papers were found. From the Google Scholar search and reviewing the citations and references of papers found, 146 papers were found.

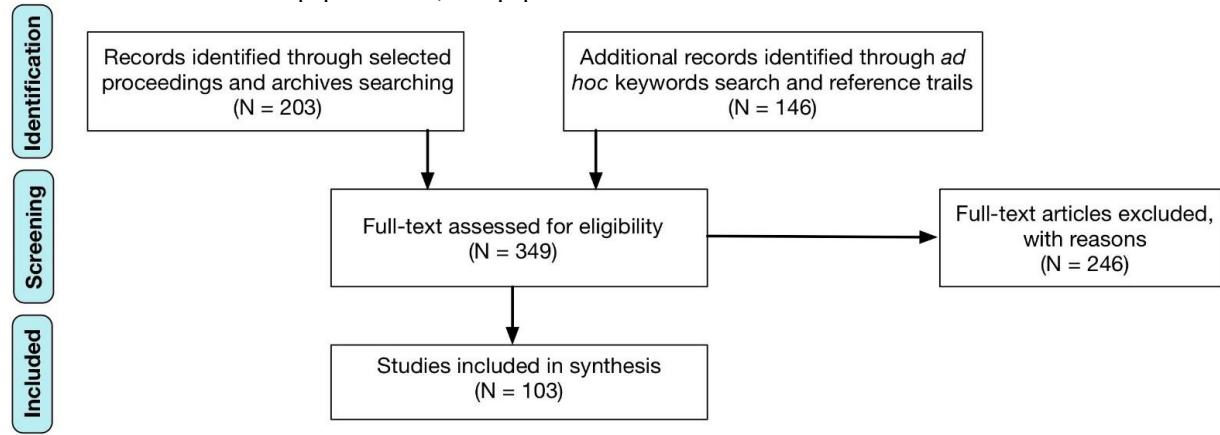

Fig. 2. Flow diagram of paper identification and selection process

*B. Paper Selection*

Out of the total 349 papers, 246 papers were excluded based on the following criteria: (i) non-relevance to wearables and education. A paper had to be relevant to both wearable technologies and education or learning to be included; and (ii) papers that did not include sufficient content for us to determine the overall scope of the work. These were often abstracts or poster papers. After the paper selection process, a total of 103 papers were kept for inclusion in the review.

Table I lists the venues where we found the most number of relevant papers. Other conferences include conferences where only single relevant papers were found, and similarly for other journals.

Table I
Number of papers included in review from selected conferences and journals proceedings

| Venue Name | No. of Papers |
|---|---|
| ACM International Conference on Human Factors in Computing Systems (CHI) | 12 |
| IEEE International Symposium on Wearable Computers (ISWC) | 9 |
| ACM International Joint Conference on Pervasive and Ubiquitous computing (UbiComp) | 8 |
| International Conference on Learning Analytics and Knowledge (LAK) | 6 |
| Computers and Education | 5 |
| IEEE Transactions on Learning Technologies | 5 |
| ACM Special Interest Group on Computer Science Education Technical Symposium (SIGCSE) | 4 |
| Journal of Computer-Assisted Learning | 2 |
| ACM International Conference on Human-Computer Interaction with Mobile Devices and Services (Mobile HCI) | 2 |
| IEEE International Symposium for Design and Technology in Electronic Packaging | 2 |
| Theses | 4 |
| Other Conferences | 22 |
| Other Journals | 22 |
| Total | 103 |

## 4. Analysis Process

Details of all the papers were extracted into a spreadsheet. These included author names, paper titles, year of publication, keywords listed, and publication venue. Basic descriptive statistics were ran on the paper details, such as calculating the number of papers published per year. All papers were then analyzed to identify the following:

- Paper type (the main contributions of the paper addressed);
- Paper description (a short description of the main topic of the paper);
- Paper findings (a short description of the main findings of the paper);
- Type of wearable (the type of wearable(s) addressed in the paper);
- Relationship of wearable and learning (description of how the wearable(s) was used for learning);
- Subject of learning (the subject(s) that the learning addressed, if mentioned);
- Research type (if the paper was an empirical study, whether the paper used a qualitative, quantitative, or mixed methodology);
- Population details (details about the target population addressed - age, race, gender, sample size); and
- Study setting (if relevant, the setting or physical context in which the work was conducted).

The analysis for the above fields was first done by 2 coders, who developed an initial coding scheme for appropriate fields. All the papers were distributed among 5 coders (including the 2 initial coders) in batches of 10 papers. After each batch was analyzed, the group of 5 coders met as a team to discuss any uncertainties in interpretation and codes that were potentially problematic, and the coding schemes for relevant fields were updated. After all the papers were analyzed for the fields listed above, 2 coders performed a thematic coding process on the 'relationship of wearable and learning' field for all the papers. The 2 coders performed an initial coding pass independently, and then met together to discuss the codes that each obtained. A final coding scheme was settled on, and one of the 2 coders did a second coding pass on all the papers using the coding scheme.

## 5. Findings

We first present our analysis findings answering our research question, and then present different perspectives on the findings based some of the more interesting dimensions extracted from the data.

*A. How Wearables Have Been Used for Learning*
We found 7 main ways in which wearables were used or proposed for use to support learning in the papers reviewed. Figure 3 shows the numbers of papers in each of the categories. We describe each category of use below, with example papers. Table A in the Appendix provides a complete list of all papers with references that were classified in each category. We note that papers were allowed to be classified in more than one category if they addressed more than one manner of wearable use.

The most common category was the use of wearables to **guide the structure of learning**. These papers propose using wearables to create some sort of framework to guide a learning activity, or to help students through the procedures of a learning task. For example, in the work by Arroyo et al. (2017), smartwatches or smartphones strapped as armbands were used to provide instructions to students as they engaged in multiplayer embodied games aimed to help them learn Math concepts (e.g., number sense). The smartwatch instructions guided the students to keep the intended pace of the games and provided feedback and support for individual players. Lukowicz et al. (2015) developed a *GoogleGlass* app for step-by-step guidance of students through the process of a science experiment (determining the relationship between sound frequency and the amount of water in a glass). The *GoogleGlass* also assisted the students in interpreting the results of their experiment through image processing of the video stream obtained through the smart glasses.

The second category of wearable use is to help **capture data to inform learning**. In those cases, wearables are used to collect data in some form from users performing or engaging in a learning activity. The data is used either to inform the design of the learning activity in real-time, to allow for evaluation of the user performance by the users themselves or by researchers later, or to allow the users (students or teachers) to review their learning. For example, Grünerbl, Pirkl, Weal, Gobbi, and Lukowicz (2015) provided nurses with *GoogleGlasses* coupled with smartwatches and smartphones while they engaged in a simulated CPR training exercise. The smart glasses captured a variety of data such as head movement and orientation, while the smartwatch captured hand-related motions. The authors proposed that the data could be used both to answer interesting research questions about emergency training situations, as well as to provide feedback to the nurses about their performance. In the work by Scholl, Wille, and Van Laerhoven (2015), a *GoogleGlass* and a smartwatch allowed students performing biology wet lab experiments to both automatically (e.g., by recording procedures performed through motion capture) and manually take notes (e.g., triggering a video or photo capture through the *GoogleGlass*) about their experiments.

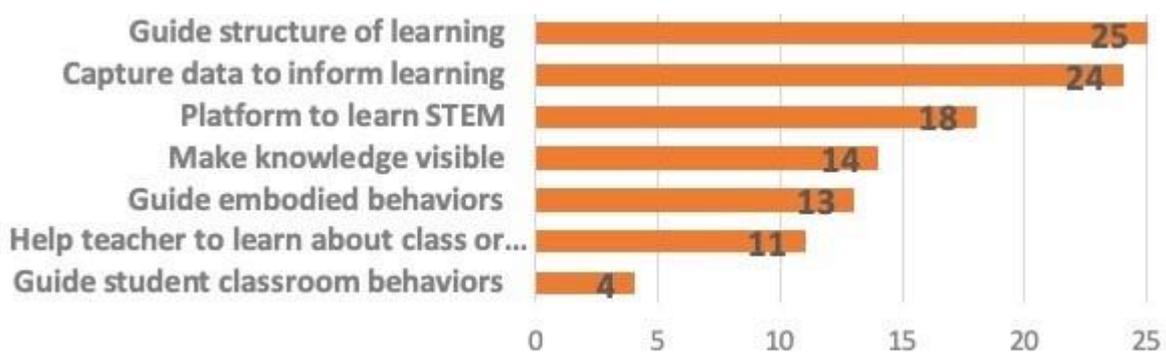

Fig. 3. Number of papers in coded categories of how wearables have been used for learning. Note: 4 papers were classified into 2 categories

The third most prevalent category was the use of wearables as a **platform to learn STEM** (Science, Technology, Engineering, Math). Many papers in this category address computer science or general engineering education, such as the learning of programming. These papers positioned wearables as a target platform or device for students to write code for. Common motivations for doing so include the proposition that students tend to express greater interest in wearables as opposed to other more typical platforms, especially in lower grades, and the fact that wearables possess many hardware limitations (e.g., processing speed, memory management) that students, especially at higher grades, should learn to handle in their computing education. For instance, Esakia, Niu, and McCrickard (2015) proposed a curriculum for a mobile development course for undergraduate students centered on programming for smartwatches. A large number of work in this category engage students in the development of e-textiles. For example, Lau, Ngai, Chan, and Cheung (2009) designed and organized a programming course that focused on wearables in fashion and design for middle school students. Students built lighted patterns on tshirts and embedded different kinds of sensors on textiles. The authors reported that their course was partially successful at increasing the students' interest in science and computing.

The category of **making knowledge visible** is an interesting one. Here, wearables are used to support learning by making explicit and/or visible abstract concepts or information that typically has no physical representation. For instance, Norooz, Mauriello, Jorgensen, McNally, and Froehlich (2015) created tshirts with sewn-on designs of the human anatomy parts (liver, heart, intestine, etc.). The tshirts were embedded with electronic LED circuits to highlight different aspects of the internal organs. Their study showed that by wearing and interacting with the tshirts, students aged 6 to 12 learned more about the sizes, positions, and functions of different human organs. Making knowledge explicit does not necessarily need to be through visual means. Pataranutaporn, Vega Gálvez, Yoo, Chhetri, and Maes (2020) rendered one's mentors' collected words of wisdom explicit through vocalization delivered through smart glasses based on the user's detected context.

The category of **guiding embodied behaviors** includes papers that make use of wearables to support learning through embodied processes such as haptic feedback. Johnson, van der Linden, and Rogers (2010), for example, developed a wearable motion capture jacket that provides vibrotactile feedback to guide users in adopting the correct position when learning to play the violin. Similarly, Huang, Do, and Starner (2008) developed gloves that provide vibrations on the users' fingers corresponding to the notes that need to be played for a particular song on the piano.

In the category **of helping teachers to learn about the class or the students**, wearables are designed to support teachers in the different tasks of teaching. The two most common tasks addressed were to help in classroom management and to help teachers understand the status of their students. Papers in this category often include analytics dashboards or wearable notifications. An example of a paper in this category is Quintana et al.'s (Quintana, Quintana, Madeira, and Slotta, 2016) in which they presented findings from co-design sessions with teachers about possible uses of wearable technologies and implemented a prototype smartwatch application. Some examples of the smartwatch app for teachers including sending reminders about participating, scheduling and logistical arrangements for a particular lesson, and real-time notifications about students' submitted assignments.

The category of guiding student classroom behaviors addresses the use of wearables to help students regulate their own behaviors in learning environments. For example, Zheng and Genaro Motti (2018) designed, developed, and tested a smartwatch application that provides different types of notifications to students with intellectual and developmental disabilities to help them integrate in a regular classroom setting. A notification on the smartwatch app, for instance, reminds the students to raise their hands to talk, or to moderate their voice volume when they speak.

*B. Perspectives on Ways of Wearable Use for Learning*

1)Historical Perspective: Figure 4 shows the distribution of all papers related to wearables for learning over the years, ranging from 2002 to 2020. Research on the use of wearables for learning purposes began to rapidly increase as from 2014, peaking in 2015 and 2016. This is probably due to the public release of the GoogleGlass in 2014 and the Apple Watch in 2015, both accompanied with much hype and bringing the idea of wearables to the forefront of public imagination. Figure 5 shows the breakdown of the papers by the 7 ways of wearable use for learning that we identified. The use of wearables to capture data to inform learning was one of the early uses of wearables that was researched, followed by using wearables as an accessible and motivating platform to help students learn programming. During the peak of research on the topic, research focused on the use of wearables to guide the structure of learning, and in recent years, we see a return to intensive research on wearable use to capture data for learning.

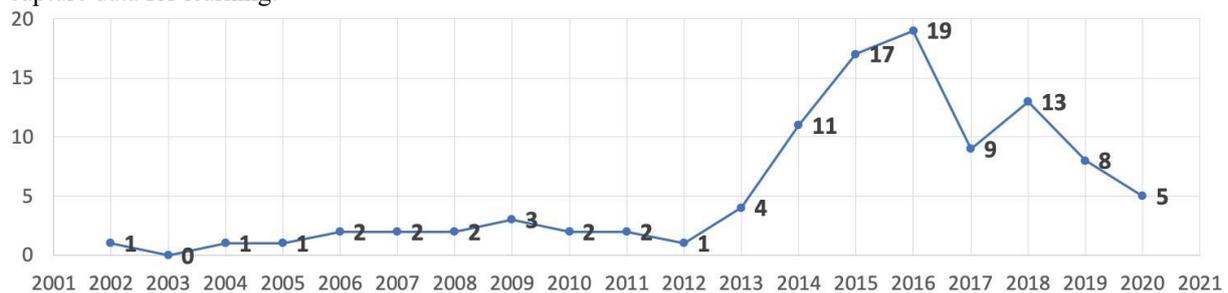

Fig. 4. Number of papers on wearables for learning by year of publication

2)Types of Wearables: In terms of the types of wearables that the reviewed papers involved, Figure 6 shows that much research focused on the use of **smart glasses or headsets** (e.g., Kawai, Mitsuhara, and Shishibori (2015); Kumar, Krishna, Pagadala, and Kumar (2018); Ponce, Menendez, Oladeji, Fryberger, and Dantuluri (2014)), with the bulk of the papers using them to guide the structure of student learning. Smart glasses are followed by **smartwatches** (e.g., Shadiev, Hwang, and Liu (2018); Zheng and Genaro Motti (2018)), which have been used for a diversity of purposes across the 7 ways of wearable use. **Smart clothing** (e.g., Norooz et al. (2015); Reichel, Schelhowe, and Grüter (2006)) is dominated by its use as a platform to learn programming, especially in the form of e-textiles. Custom wearable devices were also built in some research, such as in the work by Ryokai, Su, Kim, and Rollins (2014). **Smart wristbands**, which differ from smartwatches in that they do not possess a screen display, commonly consisted of fitness trackers, such as the FitBits used in Lee, Drake, and Thayne (2016). Researchers sometimes directly instrumented participants with **wearable sensors** (e.g., Prieto, Sharma, Dillenbourg, and Jesús (2016)), or designed **gloves** instrumented with sensors (e.g., Myllykoski, Tuuri, Viirret, and Louhivuori (2015)). Others only addressed wearables in general without explicitly referencing any specific types (e.g., Labus, Milutinovic, Stepanic, Stevanovic, and Milinovic (2015)). Last but not least, some papers built wearables specifically in the form of **badges** (e.g., Park et al. (2002); Watanabe, Matsuda, and Yano (2013)).

| | Capture data | Guide embodiment | Guide structure | Guide behaviors | Help teacher | Make knowledge visible | For programming |
|---|---|---|---|---|---|---|---|
| 2002 | 2 | 0 | 0 | 0 | 0 | 0 | 0 |
| 2004 | 1 | 0 | 0 | 0 | 0 | 0 | 0 |
| 2005 | 1 | 0 | 0 | 0 | 0 | 0 | 0 |
| 2006 | 0 | 0 | 0 | 0 | 0 | 0 | 2 |
| 2007 | 1 | 0 | 0 | 0 | 0 | 0 | 1 |
| 2008 | 0 | 1 | 0 | 0 | 0 | 0 | 1 |
| 2009 | 0 | 0 | 0 | 0 | 0 | 0 | 3 |
| 2010 | 0 | 1 | 0 | 0 | 0 | 0 | 1 |
| 2011 | 0 | 0 | 0 | 0 | 0 | 0 | 2 |
| 2012 | 0 | 1 | 0 | 0 | 0 | 0 | 0 |
| 2013 | 0 | 1 | 1 | 1 | 0 | 1 | 0 |
| 2014 | 4 | 3 | 2 | 0 | 2 | 2 | 0 |
| 2015 | 4 | 2 | 6 | 0 | 0 | 4 | 2 |
| 2016 | 5 | 0 | 5 | 0 | 4 | 4 | 2 |
| 2017 | 2 | 0 | 3 | 1 | 0 | 0 | 3 |
| 2018 | 1 | 2 | 6 | 1 | 4 | 0 | 0 |
| 2019 | 6 | 1 | 1 | 0 | 0 | 1 | 0 |
| 2020 | 1 | 0 | 0 | 0 | 1 | 2 | 1 |

Fig. 5. Ways of use for learning by year of publication

| | Capture data | Guide embodiment | Guide structure | Guide behaviors | Help teacher | Make knowledge visible | For programming | Total |
|---|---|---|---|---|---|---|---|---|
| Badge | 2 | 0 | 0 | 1 | 1 | 0 | 0 | 4 |
| Glasses | 7 | 1 | 17 | 0 | 4 | 5 | 0 | 34 |
| Smart wristband | 8 | 1 | 2 | 0 | 1 | 1 | 0 | 13 |
| Smartwatch | 5 | 1 | 3 | 2 | 6 | 1 | 2 | 20 |
| Gloves | 0 | 4 | 0 | 0 | 0 | 0 | 1 | 5 |
| Clothing | 0 | 5 | 1 | 0 | 0 | 3 | 11 | 20 |
| Custom device | 3 | 3 | 1 | 0 | 1 | 2 | 4 | 14 |
| Sensors | 2 | 0 | 0 | 0 | 2 | 0 | 0 | 4 |
| General | 2 | 0 | 1 | 0 | 0 | 2 | 0 | 5 |

Fig. 6. Wearable types by ways of use for learning

3) *Target Populations:* We analyzed the reviewed papers for target populations addressed with respect to the use of wearables for learning. We first categorized broadly the population types into students or teachers. The bulk of the papers addressed only students (82 papers or 79.6%). Eleven papers (10.7%) addressed only teachers, and 12 other papers addressed both students and teachers (11.7%). Among the papers that addressed students (including those addressing both students and teachers), we coded the age levels of the students. The following scheme was used for coding: (i) < 6 years old or < Grade 1 was coded as **PreK**; (ii) 6 to 12 years old or Grades 1 to 5 was coded using the term Elementary school; (iii) 14 to 15 years old or Grades 6 to 9 was coded as **Middle school**; (iv) 16 to 18 years old or Grades 10 to 12 was coded as **High school**; (v) 19 to 22 years old was coded as **Undergraduates;** (vi) 23 to 50 years old was coded as Adults; and (vii) > 50 years old was coded as **Older adults**. References to non-US school systems such as 'primary school' were appropriately converted. When neither the average age, age ranges or grade levels of the students were mentioned in the paper, conjectures as to an appropriate level were made based on the complexity of the topic being addressed, if possible. For example, a paper addressing the study of gravitational physics is likely to target undergraduate students, even if the population age range was not explicitly specified. If an informed conjecture was possible, the target population was coded as **Not specified**.

Figure 7 shows the distribution of papers by target populations and ways of wearable use. Some trends are made evident. Most noticeably, while wearable use to capture data to inform learning and to guide learning structure have been explored across the range of age levels, research on other ways of learning are concentrated at some age levels. For example, wearables as a platform to learn programming tend to be used mostly at the elementary school-aged level, and wearables to make knowledge visible are applied mostly at the elementary and undergraduate levels.

|  | Capture data | Guide embodiment | Guide structure | Guide behaviors | Help teacher | Make knowledge visible | For programming | Total |
|---|---|---|---|---|---|---|---|---|
| PreK | 2 | 0 | 2 | 0 | 1 | 1 | 2 | 8 |
| Elementary | 6 | 0 | 5 | 1 | 1 | 7 | 9 | 29 |
| Middle | 2 | 0 | 4 | 1 | 0 | 0 | 7 | 14 |
| High | 2 | 0 | 1 | 1 | 0 | 0 | 4 | 8 |
| Undergraduate | 7 | 3 | 7 | 1 | 2 | 6 | 5 | 31 |
| Adults | 5 | 9 | 5 | 1 | 0 | 2 | 2 | 24 |
| Older adults | 3 | 0 | 0 | 0 | 0 | 1 | 1 | 5 |
| Not specified | 4 | 3 | 3 | 0 | 0 | 0 | 2 | 12 |

Fig. 7. Target populations by ways of use for learning Note: Papers were classified in multiple categories if they addressed more than one population

|  | Capture data | Guide embodiment | Guide structure | Guide behaviors | Help teacher | Make knowledge visible | For programming | Total |
|---|---|---|---|---|---|---|---|---|
| Anywhere | 4 | 0 | 1 | 0 | 0 | 0 | 0 | 5 |
| Lab settings | 4 | 6 | 7 | 0 | 3 | 1 | 2 | 23 |
| Workshop | 1 | 0 | 0 | 0 | 0 | 0 | 5 | 6 |
| Formal settings | 13 | 1 | 9 | 3 | 5 | 5 | 6 | 42 |
| Semi-formal settings | 0 | 0 | 4 | 0 | 0 | 4 | 2 | 10 |
| Informal settings | 0 | 2 | 0 | 0 | 0 | 0 | 0 | 2 |
| None | 8 | 3 | 4 | 0 | 3 | 6 | 4 | 28 |

Fig. 8. Learning settings by ways of wearable use. Note: Papers were classified in multiple categories if they involved more than one setting type

*4) Learning Settings:* Looking at the types of settings that the papers on wearables for learning addressed, we found that most papers involved formal settings, followed by lab-based settings and semi-formal settings. Informal settings, together with the conduct of workshops, were less common. No setting was specified or could be identified in 28 of the papers. In our classification, **formal settings** consisted of mostly school or classroom environments (e.g., Quintana et al. (2016)), and sometimes, learning centers (e.g., Teeters (2007)). **Lab settings** were constrained, controlled environments typically in research labs (e.g., Russell et al. (2014)). The category of **semi-formal settings** (e.g., Kazemitabaar et al. (2017); Leue, Jung, and tom Dieck (2015)), included afterschool programs, summer camps, museums, libraries and art galleries. Semi-formal settings were characterized by the presence of some sort of structure to guide learning, although the rigidity of that structure varied and was often flexible. The category **Anywhere** was when wearables could be used across a variety of settings, or anywhere desired (e.g., at home, in vehicle, etc.). **Workshops** entailed researcher-organized sessions where the activities are predetermined (e.g., Kuznetsov et al. (2011)). **Informal settings** were settings that were informal in the context of learning, i.e., where learning is not the main goal and could happen incidentally or in an unstructured manner (e.g., surgical room (Knight, Gajendragadkar, and Bokhari (2015); Moshtaghi et al. (2015); Ponce et al. (2014)), dance hall (Hallam, Keen, Lee, McKenna, and Gupta, 2014), indoor ski resort (Spelmezan, 2012)).
Figure 8 shows how these setting types are distributed across the 7 ways of wearable use. Of note, using wearables to guide embodied behaviors has mostly been studied in lab settings, and research is scarce in informal settings and unconstrained, everyday environments (anywhere).

Table II
Summary of review findings

| Aspect | Findings |
|---|---|
| Ways of wearable use for learning | 7 different ways can be identified of how wearables have been used to support learning. See Figure 3. |
| Historical view | Research on wearables for learning accelerated as from 2013, peaked in 2015-2016, and seems to be on a downward trend since then. |
| Wearable types | A diverse distribution across wearable types, with emphasis on smart glasses, can be seen. |
| Target users | Users addressed are mostly undergraduate aged or elementary school-aged students. |
| Learning settings | Research is predominantly conducted in formal learning settings. |

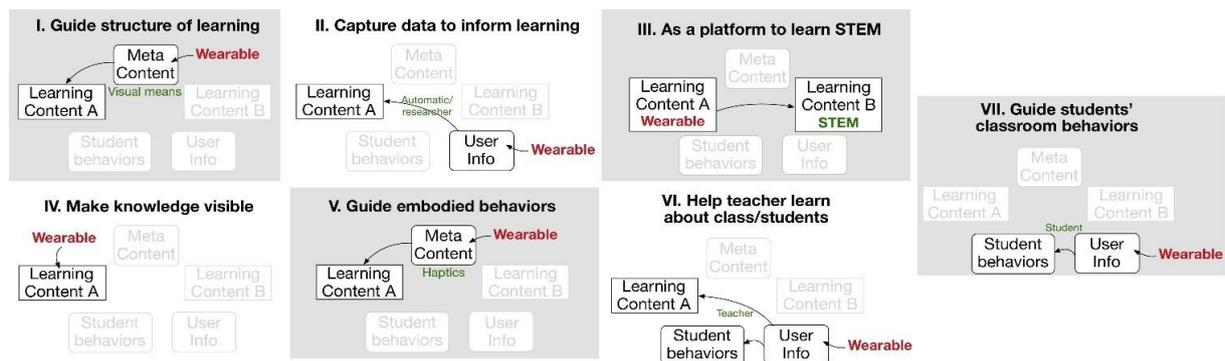

Fig. 9. Design space based on the 7 identified ways of wearable use for learning

## 6. Discussion

We conducted a systematic survey of research that has been carried out thus far on wearables for learning or for educational purposes. Our key findings are summarized in Table II. We propose a framework that can help to make sense of the overall design space of wearable use for learning
Our proposed framework consists of 5 main components that wearables for learning research: **Learning content A** (essentially, the target content to be learned - the subject area); **Learning content B** (other content to be learned that is not the main target content - sub-topic); **Meta content** (content that frames the content to be learned instructions, guides, context information, etc.); **User info** (data or information about the user); and **Student behaviors** (how students behave in the learning context either generally or with respect to specific behaviors). The 7 manners of use address different combinations of the 5 components in different ways, as illustrated in Figure 9. The framework thus also helps to make explicit what components have not been combined so far, and the unexplored design space.
The predominant use of wearables is to 'guide the structure of a learning activity' (I. in Figure 9) by providing Meta Content, such as instructions, prompts or frameworks, mainly using visual means (text and graphics) to assist the learning of Learning Content A. This manner of use is similar to 'guiding embodied behaviors' (V), but in the latter case, the guidance is provided through haptics instead of visual means. 'Capturing data to inform learning' (II) collects User Information, such as physiological data, through wearables, and uses that information to adjust Learning Content A, either through automatic methods or through researcher intervention. In the use of wearables as a 'platform to learn STEM' (III), an additional topic to be learned (Learning Content B) is added.

Students learn about wearables, and through doing so, learn about the main content (Learning Content A), which is typically a STEM subject most often programming or computer science. Using wearables to 'make knowledge visible' (IV) involves only one component of the framework. Wearables are directly used to make Learning Content A explicit to students. In the manner of use of 'helping teachers to learn about their classes or students' (VI), wearables collect and provide User Information (e.g., attention level) to the teacher, who is responsible to adjust the learning content or intervene using whatever means he/she sees fit. And in the last manner of use of 'guiding students' classroom behaviors' (VII), wearables collect and provide User Information (e.g., class participation level) to students themselves, who may regulate their own behaviors in the learning context.

We now discuss findings on the different ways of wearable use for learning with respect to the various dimensions analyzed. From a **historical perspective**, the research on wearables for learning peaked in 2015 and 2016. It is highly likely that this peak was due to the release of accessible wearables, such as the Google Glass. This shows how research can be driven by technological innovation. However, there was a rapid decrease in research after the peak in 2016. A possible explanation may be that development in terms of the capabilities of wearables flattened out after 2016. After all, the GoogleGlass was quickly retired from public access, and the Apple Smartwatch was immediately touted more as a general healthcare-focused device. On another note, it is also possible that no strong rationale has yet to be developed in research to justify why wearables should be used for learning. Yet, the space of educational wearables is less than fully explored, as our review results show. The use of wearables to capture data to inform learning seems currently to be opportunities to explore further across the other ways of wearable use we identified. In the discussion of our review findings below, we highlight potential open areas where future research may be needed, as summarized in Table III.

Table III
Summary of proposed open research directions

| Motivations | Possible Directions |
| --- | --- |
| Expanding the scope of wearables use | Research exploring different combinations of learning content A, learning content B, meta content, user info, and student behaviors for how wearables can be used to support learning. |
| Need for more theoretical rationale | Development of theoretical arguments for the use of wearable for the purpose of learning. |
| Expanding research on smart glasses and smart clothing | Studies exploring the use of smart glasses/headsets in authentic real-world contexts; and the use of smart clothing for purposes other than as a platform to learn STEM. |
| Diversifying target users and learning settings | Research on pervasive learning across the lifespan and across a diversity of contexts. |

In terms of **types of wearables**, the most addressed was smart glasses or headsets, particularly for the purpose of guiding learning structure. Although neither glasses nor headsets are commonly used in practice for learning currently, research on these wearable types dominate. A reason could be the exciting interaction possibilities that such form factor offers such as augmented reality (AR) and speech-based interaction. Further, the release of virtual reality hardware such as the *Oculus Quest* and the *Valve Index,* now allow easy development of AR applications, perpetuating the research emphasis on smart glasses and headsets. It would be interesting to see, whether and how to translate the research with smart glasses to authentic real-world uses. Another point of interest is with regards to smart clothing research, which primarily addresses the engagement of students in the development of e-textiles to support learning STEM subjects such as computer programming or electronics. Other exploration of smart cloths includes the use of smart clothing to guide embodied behaviors, but guiding students' classroom behaviors and helping the teacher in instruction are unexplored uses.

In terms of **target populations**, research on wearables for learning has mostly targeted college-aged populations. This is not surprising since college students are the most convenient sample for many research fields. Wearables for learning research has also significantly involved students at the elementary school level, focusing on the transition phase of cognitive development in the 9 to 12 age range, especially to assist in learning programming and STEM. Less research has targeted users at the high school level surprisingly, and few projects address learning for older adults. We see thus the potential for wearables for learning research to address lifelong learning that occurs across one's lifespan, to support general cognitive health as it applies to older people.

Finally, with respect to **learning settings**, most of the studies are done in formal settings that provide an existing structure to the learning process and predetermined activities, especially for the purpose of capturing data through wearables. An under-addressed aspect of wearable explorations is pervasive learning, i.e., learning that can happen anytime and anywhere. Certainly, this comes with more challenges (both technical and pragmatic) given the uncontrolled environments that this type of learning suggests but may be possible with technological advances in areas such as machine learning and data science. Such research could be more rewarding, and perhaps lead to a stronger rationale that appears to be currently lacking to catalyze research on wearables for learning further.

## 7. Conclusion and Limitations

This paper presented a systematic survey of research on wearable technologies for the purpose of learning. Designing for learning results in very different requirements than designing for health management. After all, cognitive advancement is as critical as physical health. Through a systematic review, we have identified specific ways in which wearables have been used to support learning so far in the literature, and proposed a framework identifying the main components addressed, such that future research directions are more evident. Our hope is that this review will help to accelerate research on wearables for learning in terms of developing suitable theoretical foundations, new wearable designs, new implementation techniques, and more refined evaluation studies.

The work in this paper has the following limitations: in terms of paper selection, papers that utilize wearables as a small part of a larger system for educational purposes may not have been included in the review. Our paper selection process only covered research where an emphasis was made on wearables. We also recognize that learning can be conceptualized in many ways. We did not identify nuances in how the various papers understood learning, but towards a view of being more inclusive, we included any conceptualization of learning, from the more ambiguous to the more specific, in our review.

## 8. Acknowledgments

This project was supported by NSF (National Science Foundation) grant #1566469, Lived Science Narratives: Meaningful Elementary Science through Wearable Technologies and NSF grant #1942937 Bridging Formal and Everyday Learning through Wearable Technologies: Towards a Connected Learning Paradigm.

Wearable Technologies for Learning

## APPENDIX

*Table A. Papers on Wearables for Learning Classified by Manner of Use*

| | **Manner of Wearable Use:** *Capture data to inform learning* | | | | | | | |
|---|---|---|---|---|---|---|---|---|
| # | Authors [ref] | Paper Title | Year | Paper Type | Wearable Types | Subject Areas | Population Type | Settings |
| 1 | Buchem et al. [52] | Designing for User Engagement in Wearable-technology Enhanced Learning for Healthy Ageing | 2015 | Design | Smart wristband | Health | Students | None |
| 2 | Buchem et al. [53] | Gamification Designs in Wearable Enhanced Learning for Healthy Ageing | 2015 | Design | Smart wristband | Health | Students | Anywhere |
| 3 | [Chu et al.] [54] | Toward Wearable App Design for Children's In-the-World Science Inquiry | 2017 | Study | Smartwatch | Science | Students | Lab, Anywhere |
| 4 | Ciolacu et al. [55] | Education 4.0 - Jump to Innovation with IoT in Higher Education | 2019 | Study | Smartwatch | Health | Students | Anywhere |
| 5 | Ciolacu et al. [56] | Enabling IoT in Education 4.0 with BioSensors from Wearables and Artificial Intelligence | 2019 | Study | Smartwatch | Math, General | Students | School |
| 6 | Coffman and Klinger [57] | Google Glass: Using Wearable Technologies to Enhance Teaching and Learning | 2015 | Study | Smart glasses | Educational psychology, Organizational behavior | Students + teachers | School |
| 7 | Ezenwoke et al. [58] | Wearable Technology: Opportunities and Challenges for Teaching and Learning in Higher Education in Developing Countries | 2016 | Study | Smart glasses | Accounting | Teachers | School |
| 8 | [Garcia et al.] [59] | Wearables for Learning: Examining the Smartwatch as a Tool for Situated Science Reflection | 2018 | Study | Smartwatch | Science | Students | Anywhere, School |
| 9 | Giannakos et al. [60] | Monitoring Children's Learning Through Wearable Eye-Tracking: The Case of a Making-Based Coding Activity | 2019 | Study | Smart glasses | Programming; Electronics | Students | Workshop |
| 10 | Giannakos et al. [61] | Fitbit for learning: Towards capturing the learning experience using wearable sensing | 2020 | Study | Smart wristband | Software engineering | Students | School |
| 11 | Grunerbl et al. [21] | Monitoring and enhancing nurse emergency training with wearable devices | 2015 | Design | Smartwatch | Health | Students | None |
| 12 | Ishimaru et al. [62] | Towards an intelligent textbook: eye gaze-based attention extraction on materials for learning and instruction in physics | 2016 | Study | Smart glasses | Physics | Students | School |
| 13 | Lu et al. [63] | Harnessing Commodity Wearable Devices to Capture Learner Engagement | 2019 | System | Smart wristband | General | Students + teachers | School |
| 14 | Di Mitri et al. [64] | Learning Pulse: Using Wearable Biosensors and Learning Analytics to Investigate and Predict Learning Success in Self-regulated Learning | 2016 | Study | Sensors | Learning | Students | None |
| 15 | Park et al. [42] | Design of a wearable sensor badge for smart kindergarten | 2002 | System | Badge | General | Students | None |
| 16 | Pijeira-Diaz et al. [65] | Investigating collaborative learning success with physiological coupling indices based on electrodermal activity | 2016 | Study | Smart wristband | Science | Students | Lab set up as classroom |
| 17 | Pijeira-Diaz et al. [66] | Sympathetic arousal commonalities and arousal contagion during collaborative learning: How attuned are triad members? | 2019 | Study | Smart wristband | Physics | Students | School |
| 18 | Prieto et al. [38] | Teaching Analytics: Towards Automatic Extraction of Orchestration Graphs Using Wearable Sensors | 2016 | Study | Sensors, Smart glasses | Math | Teachers | Lab |
| 19 | Russell et al. [44] | First "Glass" Education: Telementored Cardiac Ultrasonography Using Google Glass- A Pilot Study | 2014 | Study | Smart glasses | Medicine | Students | Lab |
| 20 | Scholl et al. [22] | Wearable digitization of life science experiments | 2014 | Design | Smart glasses | Science | Students | None |
| 21 | Spann and Schaeffer [68] | Expanding the scope of learning analytics data: preliminary findings on attention and self-regulation using wearable technology | 2017 | Study | Smart wristband | Cybersecurity | Students | Learning center |
| 22 | Steele and Steele [69] | Applying affective computing techniques to the field of special education | 2014 | Theory | General | Writing | Students | None |
| 23 | Sung et al. [70] | MIT.EDU: M-learning Applications for Classroom Settings | 2004 | System | Custom device | Finance, Business, Digital Anthropology | Students + teachers | School |
| 24 | Sung et al. [71] | Mobile-IT Education (MIT.EDU):m-learning applications for classroom settings | 2005 | System | Custom device | Finance, Business, Digital Anthropology | Students + teachers | School |
| 25 | Teeters [43] | Use of a Wearable Camera System in Conversation: Toward a Companion Tool for SocialEmotional Learning in | 2007 | Thesis | Custom device | N/A | Students | N/A |

| | | Autism | | | | | |
|---|---|---|---|---|---|---|---|
| **Manner of Wearable Use:** *Guide embodied behaviors* | | | | | | | |
| | Authors | Paper Title | Year | Paper Type | Wearable Types | Subject Areas | Population Type | Settings |
| 26 | Hallam et al. [50] | Ballet hero: building a garment for memetic embodiment in dance learning | 2014 | Design | Clothing | Dancing | Students + teachers | Dance hall |
| 27 | Huang et al. [28] | PianoTouch: A Wearable Haptic Piano Instruction System for Passive Learning of Piano Skills | 2008 | Study | Gloves | Music | Students | None |
| 28 | Johnson et al. [27] | MusicJacket: the efficacy of realtime vibrotactile feedback for learning to play the violin | 2010 | Study | Smart wristband | Music | Students | Lab |
| 29 | Kutafina et al. [72] | Wearable sensors in medical education: supporting hand hygiene training with a forearm EMG | 2015 | System | Clothing | Medicine | Students | None |
| 30 | Luzhnica et al. [73] | Passive haptic learning for vibrotactile skin reading | 2018 | Study | Gloves | Skin reading | Students | Lab |
| 31 | Matsushita and Iwase [74] | Detecting strumming action while playing guitar | 2013 | System | Custom device; Clothing | Music | Students | Lab |
| 32 | Myllykoski et al. [39] | Prototyping hand-based wearable music education technology | 2015 | System | Gloves | Music | Students + teachers | None |
| 33 | Pescara et al. [75] | Reevaluating passive haptic learning of morse code | 2019 | Study | Custom device; Clothing | Morse code; Skin reading | Students | Lab |
| 34 | Ponce et al. [32] | Emerging technology in surgical education: combining real-time augmented reality and wearable computing devices | 2014 | Theory | Smart glasses | Medicine | Students | Surgical room |
| 35 | Seim et al. [76] | Passive haptic learning of Braille typing | 2014 | Study | Gloves | Typing | Students | Lab |
| 36 | Seim et al. [77] | Towards haptic learning on a smartwatch | 2018 | Study | Smartwatch | Morse code | Students | Lab |
| 37 | Spelmezan [51] | An investigation into the use of tactile instructions in snowboarding | 2012 | Study | Custom device; Clothing | Snowboarding | Students | Indoor ski resort |
| **Manner of Wearable Use:** *Guide the structure of learning* | | | | | | | |
| | Authors | Paper Title | Year | Paper Type | Wearable Types | Subject Areas | Population Type | Settings |
| 38 | Arroyo et al. [19] | Wearable learning: Multiplayer embodied games for math | 2017 | Study | Smartwatch | Math | Students | School |
| 39 | Bower and Sturman [78] | What are the educational affordances of wearable technologies? | 2015 | Study | Smart glasses | General | Teachers | None |
| 40 | Cheng and Tsai [79] | A case study of immersive virtual field trips in an elementary classroom: Students' learning experience and teacher-student interaction behaviors | 2019 | Study | Smart glasses | Social studies | Students | School |
| 41 | Dieck et al. [80] | Enhancing art gallery visitors' learning experience using wearable augmented reality: generic learning outcomes perspective | 2018 | Study | Smart glasses | Art | Students | Art gallery |
| 42 | Engen et al. [81] | Wearable Technologies in the K-12 Classroom- Cross-disciplinary Possibilities and Privacy Pitfalls | 2018 | Study | Smart wristband | Physical education; Social studies; Math | Students | School |
| 43 | Hatami [82] | A study on students attitude toward employing smart glasses as a medium for e-learning | 2016 | Thesis | Smart glasses | Language | Students | Campus; Home; Library |
| 44 | Kawai et al. [31] | Tsunami Evacuation Drill System Using Smart Glasses | 2015 | Design | Smart glasses | Disaster education | Students | School |
| 45 | Kazemitabaar et al. [45] | MakerWear: A Tangible Approach to Interactive Wearable Creation for Children | 2017 | Study | Clothing | STEM | Students | Museum; Afterschool program |
| 46 | Kommera et al. [83] | Smart Augmented Reality Glasses in Cybersecurity and Forensic Education | 2016 | Theory | Smart glasses | Cybersecurity ; Forensics | Students | None |
| 47 | Leue et al. [46] | Google Glass Augmented Reality: Generic Learning Outcomes for Art Galleries | 2015 | Study | Smart glasses | Art | Students | Art gallery |
| 48 | Lindberg et al. [84] | Enhancing Physical Education with Exergames and Wearable Technology | 2016 | Study | Smart wristband | Physical education | Students | School |
| 49 | Liu [85] | Tangram Race Mathematical Game: Combining Wearable Technology and Traditional Games for Enhancing Mathematics Learning | 2014 | Thesis | Custom device | Math | Students | School |
| 50 | Liu and Chiang [86] | Smart glasses based intelligent trainer for factory new recruits | 2018 | System | Smart glasses | Industrial tasks | Students | Lab set up as factory |
| 51 | Lukowicz et al. [20] | Glass-physics: using google glass to support high school physics experiments | 2015 | Study | Smart glasses | Physics | Students | Lab |
| 52 | Moshtaghi et al. [48] | Using Google Glass to Solve Communication and Surgical Education Challenges in the Operating Room | 2015 | Study | Smart glasses | Medicine; Surgery | Students + teachers | Surgical room |
| 53 | Scholl et al. [22] | Wearables in the wet lab: a laboratory system for capturing and guiding experiments | 2015 | System | Smart glasses; Smartwatch | Science | Students | School |
| 54 | Shadiev et al. [34] | Study of the use of wearable devices for healthy and enjoyable English as a foreign language learning in authentic contexts | 2018 | Study | Smartwatch | Language | Students | School |
| 55 | Spitzer et al. [89] | Distance Learning and Assistance Using Smart Glasses | 2018 | Study | Smart glasses | Industrial tasks | Teachers | Lab |
| 56 | Spitzer et al. [88] | Project Based Learning: from the Idea to a Finished LEGO Technic Artifact, Assembled by Using Smart Glasses | 2017 | Study | Smart glasses | Industrial tasks | Students | Lab |

| | Authors | Paper Title | Year | Paper Type | Wearable Types | Subject Areas | Population Type | Settings |
|---|---|---|---|---|---|---|---|---|
| 57 | Spitzer et al. [87] | Use cases and architecture of an information system to integrate smart glasses in educational environments | 2016 | Theory | Smart glasses | Knitting | Students + teachers | Lab |
| 58 | Vallurupalli et al. [90] | Wearable technology to improve education and patient outcomes in a cardiology fellowship program- a feasibility study | 2013 | Study | Smart glasses | Medicine | Students | None |
| 59 | Vishkaie [91] | Can wearable technology improve children's creativity? | 2018 | Study | General | General | Students | Lab |
| 60 | Weppner et al. [92] | Physics Education with Google Glass gPhysics Experiment App | 2014 | System | Smart glasses | Physics | Students | None |
| **Manner of Wearable Use: *Guide students' classroom behaviors*** | | | | | | | | |
| | Authors | Paper Title | Year | Paper Type | Wearable Types | Subject Areas | Population Type | Settings |
| 61 | Watanabe and Yano [41] | Using wearable sensor badges to improve scholastic performance | 2013 | Study | Badge | General | Students + teachers | School |
| 62 | Zheng and Motti [30] | Assisting students with intellectual and developmental disabilities in inclusive education with smartwatches | 2018 | Design | Smartwatch | General | Students | School |
| 63 | Zheng and Motti [93] | Wearable Life: A Wrist-Worn Application to Assist Students in Special Education | 2017 | Design | Smartwatch | General | Students + teachers | Learning center |
| **Manner of Wearable Use: *Help teachers to learn about the class or the students*** | | | | | | | | |
| | Authors | Paper Title | Year | Paper Type | Wearable Types | Subject Areas | Population Type | Settings |
| 64 | de la Guia et al. [94] | Introducing IoT and wearable technologies into task-based language learning for young children | 2016 | Study | Smartwatch | Language | Students | Lab set up as classroom |
| 65 | Holstein et al. [95] | The Classroom as a Dashboard: Co-designing Wearable Cognitive Augmentation for K-12 Teachers | 2018 | Study | Smart glasses | General | Teachers | Lab |
| 66 | Kumar et al. [2] | Use of smart glasses in education - a study | 2018 | Theory | Smart glasses | General | Teachers | None |
| 67 | Llorente and Morant [96] | Wearable Computers and Big Data: Interaction Paradigms for Knowledge Building in Higher Education | 2014 | Theory | Smartwatch; Smart glasses | General | Teachers | None |
| 68 | Martinez Maldonado et al. [97] | Teacher Tracking with Integrity: What Indoor Positioning Can Reveal About Instructional Proxemics | 2020 | Study | Badge; Custom device | Design; Healthcare; Science | Teachers | School |
| 69 | Martinez Maldonado et al. [98] | Physical Learning Analytics: A Multimodal Perspective | 2018 | Design | Sensors | Dancing; Healthcare | Teachers | School |
| 70 | Pirkl et al. [99] | Any Problems? a wearable sensorbased platform for representational learning-analytics | 2016 | Study | Smartwatch | Physics | Students | Lab |
| 71 | Quintana et al. [29] | Keeping Watch: Exploring Wearable Technology Designs for K-12 Teachers | 2016 | Study | Smartwatch | Astronomy | Teachers | School |

| | Authors | Paper Title | Year | Paper Type | Wearable Types | Subject Areas | Population Type | Settings |
|---|---|---|---|---|---|---|---|---|
| 72 | Ueda and Ikeda [100] | Stimulation Methods for Students' Studies using Wearable Technology | 2016 | System | Smartwatch; Smart wristband | General | Students + teachers | School |
| **Manner of Wearable Use: *Make knowledge visible*** ||||||||||
| | Authors | Paper Title | Year | Paper Type | Wearable Types | Subject Areas | Population Type | Settings |
| 73 | Knight et al. [49] | Wearable technology: using Google Glass as a teaching tool | 2015 | Study | Smart glasses | Medicine | Teachers | Surgical room |
| 74 | Kuhn et al. [101] | gPhysics- Using Smart Glasses for Head-Centered, Context-Aware Learning in Physics Experiments | 2016 | Study | Smart glasses | Physics | Students | None |
| 75 | Labus et al. [40] | Wearable Computing in EEducation | 2015 | Design | General | General | Students | None |
| 76 | Lee et al. [37] | Appropriating Quantified Self Technologies to Support Elementary Statistical Teaching and Learning | 2016 | Study | Smart wristband | Statistics | Students | School |
| 77 | Meyer et al. [102] | Investigating the effect of pretraining when learning through immersive virtual reality and video: A media and methods experiment | 2019 | Study | Smart glasses | Science | Students | School |
| 78 | Norooz [103] | BodyVis: Body Learning Through Wearable Sensing and Visualization | 2014 | Thesis | Clothing | Anatomy | Students | Afterschool program |
| 79 | Norooz et al. [104] | BodyVis: A New Approach to Body Learning Through Wearable Sensing and Visualization | 2015 | Study | Clothing | Science | Students + teachers | Lab; Afterschool program |
| 80 | Norooz et al. [105] | "That's Your Heart!": Live Physiological Sensing and Visualization Tools for Life-Relevant and Collaborative STEM Learning | 2016 | Study | Clothing | Health | Students | Afterschool program |
| 81 | Pataranutaporn et al. [26] | Wearable Wisdom: An Intelligent Audio-Based System for Mediating Wisdom and Advice | 2020 | System | Smart glasses | General | Students | None |
| 82 | Peppler and Glosson [105] | Learning About Circuitry with Etextiles in After-School Settings | 2013 | Study | Custom device | Electronics | Students | None |
| 83 | Pham and Hwang [106] | Card-based design combined with spaced repetition: A new interface for displaying learning elements and improving active recall | 2016 | Study | Smartwatch | Language | Students | None |
| 84 | Ryokai et al. [36] | EnergyBugs: energy harvesting wearables for children | 2014 | Study | Custom device | Energy | Students | Summer camp; School |
| 85 | Thees et al. [107] | Effects of augmented reality on learning and cognitive load in university physics laboratory courses | 2020 | Study | Smart glasses | Physics | Students | School |
| **Manner of Wearable Use: *As a platform to learn STEM*** ||||||||||
| | Authors | Paper Title | Year | Paper Type | Wearable Types | Subject Areas | Population Type | Settings |
| 86 | Brady et al. [108] | All Roads Lead to Computing: Making, Participatory Simulations, and Social Computing as Pathways | 2017 | Study | Custom device | Programming | Students | School |
| | | to Computer Science | | | | | | |
| 87 | Buechley et al. [109] | Towards a curriculum for electronic textiles in the high school classroom | 2007 | Theory | Clothing | Programming; Electronics | Students | School; Workshop |
| 88 | Burg [110] | A STEM Incubator to Engage Students in Hands-on, Relevant Learning: A Report from the Field | 2016 | Study | Custom device | Programming; Electronics | Students | School |
| 89 | Eisenberg et al. [111] | Invisibility Considered Harmful: Revisiting Traditional Principles of Ubiquitous Computing in the Context of Education | 2006 | Theory | Clothing | Programming | Students | Lab |
| 90 | Esakia et al. [23] | Augmenting Undergraduate Computer Science Education With Programmable Smartwatches | 2015 | Study | Smartwatch | Programming | Students | School |
| 91 | Gregg et al. [112] | A Modern Wearable Devices Course for Computer Science Undergraduates | 2017 | Theory | Custom device | Electronics | Students | School |
| 92 | Jones et al. [113] | Wearable bits: Scaffolding creativity with a prototyping toolkit for wearable e-Textiles | 2020 | Design | Clothing | Design | Students | Workshop |
| 93 | Kuznetsov et al. [47] | Breaking boundaries: strategies for mentoring through textile and computing workshops | 2011 | Study | Clothing | Programming | Students | Workshop |
| 94 | Lau et al. [24] | Learning programming through fashion and design: a pilot summer course in wearable computing for middle school students | 2009 | Study | Clothing | Programming | Students | Workshop |
| 95 | Merkouris et al. [114] | Introducing Computer Programming to Children through Robotic and Wearable Devices | 2015 | Study | Smartwatch | Programming | Students | Lab |
| 96 | Merkouris et al. [115] | Teaching Programming in Secondary Education Through Embodied Computing Platforms: Robotics and Wearables | 2017 | Study | Custom device | Programming | Students | School |
| 97 | Ngai et al. [118] | An education-friendly construction platform for wearable computing | 2009 | Design | Clothing | Programming; Electronics | Students | Summer camp |
| 98 | Ngai et al. [117] | i*CATch: a scalable plug-n-play wearable computing framework for novices and children | 2010 | Design | Clothing | Programming; Electronics | Students | Workshop |
| 99 | Ngai et al. [116] | Deploying a Wearable Computing Platform for Computing Education | 2009 | Study | Clothing | Programming; General engineering | Students | Summer camp |
| 100 | Reichel et al. [35] | Smart Fashion and Learning about Digital Culture | 2006 | Study | Clothing | Programming | Students | None |
| 101 | Reichel et al. [119] | Eduwear: Designing Smart Textiles for Playful Learning | 2008 | Design | Clothing | Programming | Students | None |
| 102 | Reimann [120] | Shaping Interactive Media with the Sewing Machine: Smart Textile as an Artistic Context to Engage Girls in Technology and Engineering Education | 2011 | Design | Gloves | Programming | Students | None |
| 103 | Reimann and Maday [121] | Smart Textile objects and conductible ink as a context for arts based teaching and learning of computational thinking at primary school | 2016 | System | Clothing | Programming | Students | None |